\documentstyle[epsfig,12pt]{article}  
\newcommand{\beq}{\begin{equation}}
\newcommand{\eeq}{\end{equation}}
\newcommand{\bea}{\vspace{0.25cm}\begin{eqnarray}}
\newcommand{\eea}{\end{eqnarray}}

\newcommand{\r}{\mbox{{\boldmath
$\rho$}}}

\newcommand{\pb}{\mbox{{\bf
p}}}
\newcommand{\kb}{\mbox{{\bf
k}}}
\newcommand{\rb}{\mbox{{\bf
r}}}

\def\lsim{\mathrel{\rlap{\lower4pt\hbox{\hskip1pt$\sim$}}
\raise1pt\hbox{$<$}}}         %less than or approx. symbol
\def\gsim{\mathrel{\rlap{\lower4pt\hbox{\hskip1pt$\sim$}}
    \raise1pt\hbox{$>$}}}         %greater than or approx. symbol
\begin{document}
\thispagestyle{empty}
%\vspace*{2cm}
\bigskip

\begin{center}

  {\large\bf
TRANSITION RADIATION IN THE QUANTUM REGIME AS A DIFFRACTIVE PHENOMENON
\\
\vspace{1.5cm}
  }
\medskip
  {\large
  D. Schildknecht$^a$ and B.G. Zakharov$^{b,c}$}
  \bigskip

{\it
$^{a}$Fakult\"at f\"ur Physik, Universit\"at Bielefeld\\
D-33501 Bielefeld, Germany\\
$^{b}$Institut  f\"ur Kernphysik,
        Forschungszentrum J\"ulich,\\
        D-52425 J\"ulich, Germany\\
%\medskip\\
$^{c}$L.D. Landau Institute for Theoretical Physics,
        GSP-1, 117940,\\ Kosygina Str. 2, 117334 Moscow, Russia
%        \medskip\\
\vspace{2.7cm}\\}

  {\bf
  Abstract}
\end{center}
{
\baselineskip=9pt
We demonstrate that the transition photon radiation and pair creation
can be interpreted as a diffractive phenomenon in terms of the light-cone wave
functions in a way similar to the Good-Walker approach \cite{GW} 
to the diffraction dissociation. Our formulas for spectra agree with those 
obtained by Baier and
Katkov \cite{BK} within the quasiclassical operator method. However, there is
some disagreement with earlier results by Garibyan \cite{G}.
}

\vspace{1.7cm}
%\pagebreak
%-------------------------------------------------------------
\noindent {\bf 1}.
The transition radiation \cite{GF} is usually discussed within the classical
electrodynamics (for reviews, see \cite{TM,GT}).
For relativistic particles the classical approach applies when the momentum of 
the radiated photon is small as compared to the momentum of the charged particle. 
The emission of hard photons with momentum comparable to the charged particle
momentum requires a quantum treatment.
The transition radiation in quantum regime and transition pair creation
was addressed many years ago by Garibyan \cite{G}. Recently the problem was 
analyzed by Baier and Katkov \cite{BK} within the quasiclassical operator 
method. The results of Ref. \cite{BK} are in some disagreement with 
those of \cite{G}.
The major purpose of the present note is to demonstrate 
that the transition radiation (and pair creation) in quantum regime 
can be interpreted as a diffractive effect in terms
of the light-cone wave functions in the spirit
of the Good-Walker approach to the diffraction dissociation \cite{GW}.
The analysis is quite simple, nevertheless, to the best of our
knowledge this physical interpretation of the transition radiation has never been
discussed in the literature. Also, our calculations clarify the
situation with disagreement between the results of Refs. \cite{BK} and \cite{G}.
Our results confirm the results of Baier and Katkov \cite{BK} and 
disagree with Garibyan's formulas for the case of the transition radiation.

\noindent {\bf 2}.
%In quantum regime
Let us consider for definiteness the photon emission 
from a relativistic electron with energy $E_{e}\gg m_{e}$
(we use the units $\hbar=c=1$) 
which moves normally to the boundary from medium 1 to medium 2.
We choose the $z$ axis along the electron momentum and assume that the 
boundary is located at $z=0$.  
We assume that the photon momentum is sufficiently large and 
the medium effect on the photon quantum field 
may be treated in the plasma approximation \cite{Migdal,GT}. In this 
approximation the in-medium photon is described as a quasiparticle 
with nonzero mass
$m_{\gamma}=\omega_{p}$, where
$\omega_{p}=\sqrt{4\pi ne^{2}/m_{e}}$ is the plasma frequency \cite{Migdal}.
The wave functions (to leading order in the coupling constant)
of the physical electron with energy $E_{e}$ in the media 1 and 2 can be 
symbolically written as the Fock state superposition 
of the bare electron $|e\rangle$ and electron-photon state $|e'\gamma\rangle$
(we assume that in the electron-photon state $E_{\gamma}\gg m_{\gamma}$ and
$E_{e'}\gg m_{e}$) 
\beq
|e^{ph}_{i}\rangle=|e\rangle +\sum_{x,\kb}\Psi_{i}(x,\kb)|e'\gamma\rangle\,,
\label{eq:10}
\eeq
where $x=p_{\gamma,z}/p_{e,z}$ is 
the photon fractional longitudinal momentum, $\kb$ is the transverse 
momentum of the photon,  $\Psi_{i}(x,\kb)$ is the wave function 
of the electron-photon Fock state in terms of 
the variables $x$ and $\kb$. The $\Psi_{i}(x,\kb)$ is usually 
called the light-cone (or infinite-momentum frame)
wave function \cite{BKS,LB} (for a review of the light-cone formalism, 
see \cite{BPP}).
It is important that at high energies the electron-photon Fock state 
can be localized in a small region (with a size of the order of several 
units of $1/\min(E_{\gamma},E_{e'})$) which is much smaller than
the formation time for the $|e'\gamma\rangle$ state
(see below). For this reason the dynamics of the $|e'\gamma\rangle $
state in the media 1 and 2 turns out to be completely independent
of the boundary, and one may evaluate $\Psi(x,\kb)$
for infinite media neglecting  the boundary effects. 

The light-cone wave function can easily be obtained using the ordinary
perturbation formula for the first order correction to the wave function:
$|\delta \psi_{n}\rangle=|\psi_{m}\rangle 
\langle\psi_{m}|\hat{H}_{int}|\psi_{n}\rangle/(E_{n}-E_{m})$.
Using the QED interaction Hamiltonian 
$\hat{H}_{int}=e\int d\rb \bar{\psi}\hat{A}\psi$, and
writing the particle energies as 
$E\approx |p_{z}|+(m^{2}+\pb_{\perp}^2)/2|p_{z}|$
one obtains
\beq
\Psi_{i}(x,\kb)=\frac{\sqrt{x(1-x)}}{2\sqrt{\pi}}
\cdot\frac{e\bar{u_{e'}}\gamma_{\mu} u_{e} \epsilon^{\mu}}
{\kb^{2}+\kappa_{i}^{2}}\,,
\label{eq:20}
\eeq
where $\kappa^{2}_{i}=m_{e}^{2}x^{2}+m_{\gamma,i}^{2}(1-x)$,
$u_{e}$, $u_{e'}$ are the electron spinors, $\epsilon^{\mu}$ is the photon 
polarization vector (we use the normalization of the light-cone wave function
corresponding to probability of the electron-photon Fock component
given by $\frac{1}{(2\pi)^2}\int d\kb\int_{0}^{1} dx |\Psi(x,\kb)|^2$).
Note that (\ref{eq:20}) is valid when the photon and electron 
longitudinal momenta are positive. Due to large energy denominator
in the perturbative formula,
the configurations with photon (or electron) moving backward are suppressed
by a factor of the order of $\kappa_{1,2}^{2}/E_{\gamma,e'}^{2}$, and can
be neglected at high energies.

The discontinuity of the $m_{\gamma}$ on the boundary between two media 
leads to a jump in the Fock states
decomposition of the physical electron.
Evidently, the Fock component $|e'\gamma\rangle$ after passing 
through the boundary from medium 1 to 
medium 2 has the same wave function as in medium 1, i.e. $\Psi_{1}(x,\kb)$, 
which, however, does not match 
with the wave function of the $|e'\gamma\rangle$ state in medium 2. 
Thus, in terms of the physical states the projectile state after 
passing through the boundary, $|\tilde{e}_{1}\rangle$, can be written as
\beq
|\tilde{e}_{1}\rangle=|e^{ph}_{2}\rangle +\sum_{x,\kb}
[\Psi_{1}(x,\kb)-\Psi_{2}(x,\kb)]
|{e'}^{ph}_{2}\gamma^{ph}_{2}\rangle\,.
\label{eq:30}
\eeq
Evidently, the second term on the right-hand 
side of (\ref{eq:30}) describes the photon emission.  The corresponding 
spectrum 
reads
\beq
\frac{dN}{dxd\kb}=\frac{1}{(2\pi)^{2}}\left
|\Psi_{1}(x,\kb)-\Psi_{2}(x,\kb)\right|^{2}\,.
\label{eq:40}
\eeq

The above analysis of the transition radiation is 
very similar to the Good-Walker treatment  of 
the diffraction dissociation \cite{GW}. In that case the projectile wave 
function
after the target has the form $|\Psi_{f}\rangle=\hat{S}|\Psi_{in}\rangle$,
where $\hat{S}$ is the $S$-matrix, and $|\Psi_{in}\rangle$ is the wave function
of the incident particle.
In the diffraction dissociation the jump in the Fock
state decomposition of the projectile stems from the different 
scattering amplitudes for different Fock components. In the case of the 
transition radiation there is no scattering
at all. Nonetheless, similarly to the diffraction dissociation there is a 
jump in the Fock state decomposition of the projectile state,
which also leads to the inelastic process. 
Our treatment of the transition radiation in terms of the light-cone
wave functions technically is also analogues  
to the Bjorken-Kogut-Soper method \cite{BKS} for bremsstrahlung and
pair production off an external field in the high-energy limit 
based on the formulation of QED in the infinite-momentum frame \cite{KS}.

From (\ref{eq:20}) one obtains for the no electron spin flip transition
\beq
\Psi_{i}(x,\kb)=i\sqrt{\frac{\alpha}{2x}}
[\lambda_{\gamma}(2-x)+2\lambda_{e}x]\cdot\frac{\kb_{x}-i\lambda_{\gamma}\kb_{y}}
{\kb^{2}+\kappa_{i}^{2}}\,,
\label{eq:50}
\eeq
where $\alpha=1/137$, $\lambda$ denotes particle helicity 
(it is convenient to use the light-cone helicity 
basis \cite{BKS,LB}).
For the electron spin flip transition
the only nonzero component is that with $\lambda_{\gamma}=2\lambda_{e}$
\beq
\Psi_{i}(x,\kb)=\sqrt{2\alpha x^{3}}\cdot
\frac{m_{e}}
{\kb^{2}+\kappa_{i}^{2}}\,.
\label{eq:60}
\eeq
Using (\ref{eq:50}),~(\ref{eq:60}) one obtains from (\ref{eq:40}) 
\beq
\frac{dN}{dxd\kb}=\frac{\alpha}{\pi^{2}x}\left\{
\kb^{2}[1-x+\frac{x^{2}}{2}]+\frac{m_{e}^{2}x^{4}}{2}\right\}
\cdot\left[\frac{1}{\kb^{2}+\kappa_{1}^{2}}-\frac{1}{\kb^{2}+
\kappa_{2}^{2}}\right]^{2}
\,.
\label{eq:70}
\eeq
From (\ref{eq:70}) after integration over the transverse momentum 
one gets for the $x$-spectrum
\beq
\frac{dN}{dx}=\frac{\alpha}{\pi x}\left\{
[1-x+\frac{x^{2}}{2}]
\left[\frac{\kappa_{1}^{2}+\kappa_{2}^{2}}{\kappa_{2}^{2}-\kappa_{1}^{2}}
\ln{\frac{\kappa_{2}^{2}}{\kappa_{1}^{2}}}-2\right]
-\frac{m_{e}^{2}x^{4}}{2}
\left[\frac{2}{\kappa_{2}^{2}-\kappa_{1}^{2}}
\ln{\frac{\kappa_{2}^{2}}{\kappa_{1}^{2}}}-
\frac{\kappa_{1}^{2}+\kappa_{2}^{2}}{\kappa_{2}^{2}\kappa_{1}^{2}}
\right]
\right\}
\,.
\label{eq:80}
\eeq
The formulas (\ref{eq:70}),~(\ref{eq:80}) apply when $x\gg m_{\gamma,i}/E_{e}$
and $(1-x)\gg m_{e}/E_{e}$. For $x\ll 1$ our formulas agree with the 
classical results \cite{GT}.

Our analysis neglects
the backward photon emission, and in general 
bremsstrahlung with large angle between the 
photon and electron momenta $\theta_{\gamma e}\gsim 1$. This region cannot be 
discussed in 
the light-cone wave function language. However, 
the forward emission is dominated by the small angle region 
$\theta_{\gamma e}\lsim \max(m_{e}/E_{e},m_{\gamma}/E_{\gamma})$, and 
the contribution
from $\theta_{\gamma e}\sim 1$ can be neglected. In the classical approach the 
backward emission 
is much smaller than the forward one for ultrarelativistic energies \cite{GT}.
In our treatment this is a consequence of the above mentioned 
suppression of the Fock $|e' \gamma \rangle$ component
in the physical electron with negative photon longitudinal momentum.

In the above we have discussed the situation when
the electron moves normally to the boundary. 
The treatment of the transition radiation in terms of the light-cone
wave functions allows one to understand 
easily the applicability limits of the approach for non-orthogonal movement
of the initial electron. From the uncertainty relation 
$\Delta E \Delta t\gsim 1$ one can obtain
for the typical life-time of the virtual electron-photon states
in media 1 and 2 $t_{i}\sim 2E_{e}x(1-x)/\kappa_{i}^{2}$. 
If $\kappa_{1}\sim \kappa_{2}$ the quantity
$\bar{t}\sim (t_{1}+t_{2})/2$ can be viewed as a formation length, $L_{f}$, 
associated with the photon emission. 
In this case the dominating region of the photon 
transverse momentum is given by $k\sim \bar{\kappa}\sim
(\kappa_{1}+\kappa_{2})/2$. 
It corresponds 
to the typical transverse separation between photon and electron
in the photon-electron state $\rho_{\gamma e'}\sim 1/\bar{\kappa}$.
Using the above estimate one can easily obtain 
the following condition for applicability 
of the formulas obtained for normal incidence of the electron:
$\tan \psi\ll L_{f}/\rho_{\gamma e'}
\sim 2E_{\gamma}(1-E_{\gamma}/E_{e})/\bar{\kappa}$, where $\psi$ is the 
angle between the initial electron momentum and normal to the boundary.
If $\kappa_{1}$ and $\kappa_{2}$ differ strongly one should replace 
in the above inequality $\bar{\kappa}$ by $\min(\kappa_{1},\kappa_{2})$.
The above analysis shows that in the region under consideration 
$E_{\gamma}\gg \omega_{p}$ the formulas obtained for normal incidence
are valid in a broad range of the incidence angles.

In a similar way one can obtain for the cross-channel process
$\gamma\rightarrow e^{+}e^{-}$
\beq
\frac{dN}{dxd\kb}=\frac{\alpha}{2\pi^{2}}\left\{
\kb^{2}[x^{2}+(1-x)^{2}]+m_{e}^{2}\right\}
\cdot\left[\frac{1}{\kb^{2}+\kappa_{1}^{2}}-\frac{1}{\kb^{2}+\kappa_{2}^{2}}\right]^{2}
\,,
\label{eq:90}
\eeq
\bea
\frac{dN}{dx}=\frac{\alpha}{2\pi }\left\{
[x^{2}+(1-x)^{2}]
\left[\frac{\kappa_{1}^{2}+\kappa_{2}^{2}}{\kappa_{2}^{2}-\kappa_{1}^{2}}
\ln{\frac{\kappa_{2}^{2}}{\kappa_{1}^{2}}}-2\right]\right.
\left.
-m_{e}^{2}
\left[\frac{2}{\kappa_{2}^{2}-\kappa_{1}^{2}}
\ln{\frac{\kappa_{2}^{2}}{\kappa_{1}^{2}}}-
\frac{\kappa_{1}^{2}+\kappa_{2}^{2}}{\kappa_{2}^{2}\kappa_{1}^{2}}
\right]
\right\}
\,,
\label{eq:100}
\eea
where now $\kappa_{i}^{2}=m_{e}^{2}-m_{\gamma,i}^{2}x(1-x)$,
and $x=p_{e,z}/p_{\gamma,z}$. The applicability domain of (\ref{eq:90}) and
(\ref{eq:100}) is $x\gg m_{e}/E_{\gamma}$, $(1-x)\gg m_{e}/E_{\gamma}$.

Our formulas for the photon emission (\ref{eq:70}), (\ref{eq:80})
and pair production (\ref{eq:90}), (\ref{eq:100}) agree 
with those of Ref. \cite{BK}.
However, the formulas (\ref{eq:70}), (\ref{eq:80}) disagree with 
equations (13) and (14) of Ref. \cite{G} (if one rewrites them in terms of the 
variables $x$ and $\kb$). 
In the case of pair production (\ref{eq:90}) agrees with equation (18)
of Ref. \cite{G}. The limit $x\rightarrow 0$ of our $x$-spectrum (\ref{eq:100}) 
agrees with formula (19) of Ref. \cite{G} obtained for $E_{e}\ll E_{\gamma}$.

\noindent {\bf 3}.
It is instructive to see how the light-cone wave functions appear in
the derivation of the transition radiation from the ordinary perturbative
formula for the amplitude of the $e\rightarrow e'\gamma$ transition
\beq
T=-2\pi \delta(E_{e'}+E_{\gamma}-E_{e})\,
e\!\int d\rb \bar{\psi}_{e'}(\rb)\gamma_{\mu}{A^{\mu}}(\rb)\psi_{e}(\rb)\,,
\label{eq:110}
\eeq
where $\psi_{e}$, $\psi_{e'}$ are the 
the electron  wave functions,
$A^{\mu}$ is the photon wave function. Note that
Eq. (\ref{eq:110}) was the starting point of the analysis \cite{G}.

The electron wave functions (for $E_{i}\gg m_{e}$) can be written in the form
\beq
\psi_{j}(\rb)=\frac{u_{j}}{\sqrt{2E_{j}}}
\exp(iE_{j}z)\phi_{j}(z,\r)\,,
\label{eq:120}
\eeq
where $\rb=(\r,z)$, and the $\phi_{j}$ satisfies the 
Schr\"odinger equation
\beq
i\frac{\partial \phi_{j}(z,\r)}{\partial z}=
\left[-\frac{1}{2E_{j}}
\left(\frac{\partial}{\partial \r}\right)^{2}+
\frac{m_{e}^{2}}{2E_{j}}\right]\phi_{j}(z,\r)\,.
\label{eq:130}
\eeq
Evidently, one can take for the transverse electron wave functions $\phi_{j}$
\beq
\phi_{j}(z,\r)=\exp{\left\{i\left[\pb_{j}\r-z 
\frac{\pb_{j}^{2}+m_{e}^{2}}{2E_{j}}\right]\right\}}\,,
\label{eq:140}
\eeq
where $\pb_{j}$ are the transverse momenta.
The photon wave function can be written 
in a form similar to (\ref{eq:120}) (up to an obvious
change of the spin factor). Note that neglecting the terms suppressed
by factors $m_{\gamma,i}^{2}/E_{\gamma}^{2}$ one
can neglect the photon wave which propagates backward \cite{G}. Also, in this 
approximation one can use the same photon polarization vector for 
$z < 0$ and $z > 0$.
The photon transverse wave function, which satisfies an equation similar 
to (\ref{eq:130}) (but with $z$-dependent mass), can be taken in the form
\beq
\phi_{\gamma}(z,\r)=\exp{\left\{i\left[\pb_{\gamma}\r-\int_{0}^{z}d\xi 
\frac{\pb_{\gamma}^{2}+m_{\gamma}^{2}(\xi)}{2E_{\gamma}}\right]\right\}}\,,
\label{eq:150}
\eeq
where $\pb_{\gamma}$ is the photon transverse momentum, for a sharp boundary 
$m_{\gamma}(\xi<0)=m_{\gamma,1}$,
$m_{\gamma}(\xi>0)=m_{\gamma,2}$.

In terms of the transverse wave functions $\phi_{j}$ 
the transition amplitude (\ref{eq:110}) reads
\bea
T=-2\pi \delta(E_{e'}+E_{\gamma}-E_{e})
\frac{e\bar{u}_{e'}\gamma_{\mu} u_{e}\epsilon^{\mu}}
{\sqrt{8E_{e}^{3}x(1-x)}}
\cdot\int dz d\r \phi_{e'}^{*}(z,\r)
\phi_{\gamma}^{*}(z,\r)
\phi_{e}(z,\r)\,.
\label{eq:160}
\eea
After integrating over $\r$ in (\ref{eq:160}) one obtains
\beq
T=-(2\pi)^{3} \delta(E_{e'}+E_{\gamma}-E_{e})
\delta(\pb_{e'}+\pb_{\gamma}-\pb_{e})M\,,
\label{eq:170}
\eeq
\beq
M=
\frac{e\bar{u}_{e'}\gamma_{\mu} u_{e}\epsilon^{\mu}}
{\sqrt{8E_{e}^{3}x(1-x)}}
\cdot\int\limits_{\infty}^{\infty} dz
\exp{\left[-i\int_{0}^{z}d\xi 
\frac{\kb^{2}+\kappa^{2}(\xi)}{2E_{e}x(1-x)}\right]}\,,
\label{eq:180}
\eeq
where $\kb=\pb_{\gamma}(1-x)-\pb_{e'}x$,
and 
$\kappa^{2}(\xi)=m_{e}^{2}x^{2}+m_{\gamma}^{2}(\xi)(1-x)$.
The $z$-integration in (\ref{eq:180}) (for a sharp boundary between the 
two media) 
gives
\beq
M=ie
\bar{u}_{e'}\gamma_{\mu} u_{e}\epsilon^{\mu}
\sqrt{\frac{x(1-x)}{2E_{e}}}\cdot
\left(\frac{1}{\kb^{2}+\kappa_{1}^{2}}-\frac{1}{\kb^{2}+\kappa_{2}^{2}}
\right)\,.
\label{eq:190}
\eeq
The two terms in the large brackets on the right-hand side of (\ref{eq:190})
stem from the integration regions in (\ref{eq:180}) $z<0$ and $z>0$, 
respectively. 

From the Fermi golden rule one can easily obtain the spectrum
\beq 
\frac{dN}{dxd\kb}=\frac{E_{e}|M|^{2}}{(2\pi)^{3}}.
\label{eq:200}
\eeq
From (\ref{eq:20}), (\ref{eq:30}) and (\ref{eq:190}) one sees that (\ref{eq:200}) 
reproduces (\ref{eq:40}) obtained in the 
light-cone wave function approach.
The spectrum for the cross-channel process 
$\gamma\rightarrow e^{+}e^{-}$
can be obtained with the help of the replacement in $|M|^{2}$:
$|\kb|^{2}\rightarrow \kb|^{2}/x^{2}$, $x\rightarrow 1/x$. It gives
the spectrum which agrees with (\ref{eq:90}). 
%Thus, Eq. (\ref{eq:110}) leads to the same results as the analysis 
%based on the light-cone wave function approach. 
Note that the difference of the two terms on the right-hand side of (\ref{eq:190})
precisely corresponds to the jump of the light-cone wave functions
on the boundary between the media. However, now 
the nonzero amplitude $M$ (\ref{eq:190}) emerges due to
the jump in the quantity $\Delta p_{z}=p_{\gamma,z}+p_{e',z}-p_{e,z}$
when the energy is conserved,
while the nonzero difference of the light-cone wave functions
in media 1 and 2 is due to a jump in 
$\Delta E=E_{\gamma}+E_{e'}-E_{e}$ when the momentum is conserved
(since we evaluated $\Psi_{i}(x,\kb)$ for infinite media). However, 
in the high-energy limit 
$|\Delta E|=|\Delta p_{z}|$, and the two approaches are equivalent.
As was above noted Eq. (\ref{eq:110}) is the starting point of the 
analysis \cite{G}. Thus, the above mentioned disagreement of our 
results for bremsstrahlung 
(and the results of Ref. \cite{BK}) with those of Ref. 
\cite{G} is probably due to some mistakes in 
evaluating the transition matrix element in \cite{G}.

The quantum effects in the transition photon emission
become important when $x$ is not small. However, 
the transition spectrum (\ref{eq:80}) falls rapidly ($\propto 1/x^{5}$) for 
$x\gsim m_{\gamma}/m_{e}$. For the ordinary materials 
$m_{\gamma}/m_{e}\sim 10^{-5}-10^{-4}$. For this reason the radiation intensity
turns out to be small in the quantum domain $x\sim 1$. 
The quantum effects may be important for the transition radiation
in the electrosphere \cite{U1,U2,U3} of the strange stars made of 
strange quark matter
\cite{Witten} (if they exist) where the ratio $m_{\gamma}/m_{e}$ may be 
about unity 
(or even larger) \cite{JGPP,U2}.
The quantum effects are also important in the non-Abelian analog of the 
transition radiation from fast quarks/gluons traversing the 
finite-size quark-gluon plasma
produced in high-energy heavy-ion collisions \cite{Z_coherent}. 
However, in these cases, due to large densities, the induced 
radiation caused by multiple scattering \cite{Z_LPM} becomes important, and,
strictly speaking, both these mechanisms should be treated on even footing.

\noindent {\bf 4}.
In summary, we have demonstrated that the transition photon radiation 
and pair creation
may be interpreted as a diffractive phenomenon in terms of the light-cone wave
functions in a way similar to the Good-Walker approach \cite{GW} 
to the diffraction dissociation. Our results agree with those obtained 
by Baier and
Katkov \cite{BK} within the quasiclassical operator method. However,
there is some  disagreement with
Garibyan's calculations \cite{G}.

\vspace {1 cm}
\noindent
{\large\bf Acknowledgements}

\noindent
We are grateful to N.N. Nikolaev for discussions.
B.G.Z. thanks the Physical Department of the University of 
Bielefeld
and FZJ, J\"ulich, for the kind hospitality
during the time when this work was done. 
This research is supported by the grants 
DFG Schi 189/6-1 and DFG 436RUS17/101/04.

\end{document}